# Interface-induced magnetism in perovskite quantum wells

**Clayton A. Jackson and Susanne Stemmer**

Materials Department, University of California, Santa Barbara, California 93106-5050, USA

**Abstract**

We investigate the angular dependence of the magnetoresistance of thin (< 1 nm), metallic $SrTiO_3$ quantum wells epitaxially embedded in insulating, ferrimagnetic $GdTiO_3$ and insulating, antiferromagnetic $SmTiO_3$, respectively. The $SrTiO_3$ quantum wells contain a high density of mobile electrons ($\sim 7 \times 10^{14} cm^{-2}$). We show that the longitudinal and transverse magnetoresistance in the structures with $GdTiO_3$ are consistent with anisotropic magnetoresistance, and thus indicative of induced ferromagnetism in the $SrTiO_3$, rather than a nonequilibrium proximity effect. Comparison with the structures with antiferromagnetic $SmTiO_3$ shows that the properties of thin $SrTiO_3$ quantum wells can be tuned to obtain magnetic states that do not exist in the bulk material.

Proximity effects at interfaces between ferromagnetic insulators and conductors with strong spin-orbit coupling have attracted attention as components of hybrid structures for spintronics, quantum computing, and as a route to Majorana fermions [1-3]. Oxide heterostructures are particularly attractive for inducing phenomena through interfacial proximity, because the relevant phenomena, such as superconductivity, spin-orbit coupling, two-dimensional electron gases, and magnetism can all be found in a single materials class, the perovskites, allowing for high-quality epitaxial structures.

A prototypical perovskite heterostructure is that between the ferrimagnetic Mott insulator $GdTiO_3$, and the band insulator $SrTiO_3$. Such interfaces exhibit a high-density, two-dimensional electron gas (2DEG) with $\sim 3 \times 10^{14} cm^{-2}$ mobile carriers in the $SrTiO_3$ [4]. Thin (< 2 nm) quantum wells of $SrTiO_3$ embedded in $GdTiO_3$ show magnetoresistance hysteresis at low temperatures [5]. It was suggested that the proximity to the ferrimagnetic $GdTiO_3$ likely plays a role, but the magnetic state of the $SrTiO_3$ was not resolved. For example, it is possible that the $SrTiO_3$ quantum well has become magnetic due to exchange coupling. Theoretical calculations suggest a tendency towards ferromagnetism for thin quantum wells [6, 7]. In this case, the $SrTiO_3$ quantum well may exhibit magnetotransport properties typical of ferromagnets, such as anisotropic magnetoresistance (AMR). An alternative explanation is that the resistance hysteresis reflects the orientation of the magnetization in the $GdTiO_3$, without the $SrTiO_3$ quantum well itself being ferromagnetic. The latter is a non-equilibrium proximity effect that has become known as spin Hall magnetoresistance (SMR), and results from a combination of direct and inverse spin Hall effects [8-10]. Spin-orbit coupling is at the core of both SMR and AMR. As spin-related effects are important in both bulk $SrTiO_3$ [11] and $SrTiO_3$ 2DEGs [12, 13], either AMR (for a ferromagnetic quantum well) or SMR (for a non-magnetic quantum well) may occur. The two phenomena are distinguishable by the dependence of the magnetoresistance on the orientation of the magnetic field [8].

To engineer novel states at interfaces between conducting, non-magnetic perovskites and insulating magnetic perovskites, it is essential that the magnetic state of these interfaces be understood. In this Letter, we report angular-dependent magnetoresistance studies of narrow $SrTiO_3$ quantum wells, embedded in ferrimagnetic $GdTiO_3$ and antiferromagnetic $SmTiO_3$, respectively. The results are consistent with exchange coupling induced ferromagnetism in the $SrTiO_3$ in case of interfaces with $GdTiO_3$ and, more indirectly, with induced antiferromagnetism in case of $SmTiO_3$.

$GdTiO_3/SrTiO_3/GdTiO_3$ quantum well structures were grown by hybrid molecular beam epitaxy on (001) $(LaAlO_3)_{0.3}(Sr_2AlTaO_6)_{0.7}$ (LSAT) crystals. The data reported here is for a sample with 4 nm top and bottom $GdTiO_3$ layers, and 0.8 nm $SrTiO_3$ (about 3 SrO layers [14]). Samples of $SmTiO_3/SrTiO_3/SmTiO_3$ with varying $SrTiO_3$ thicknesses were also investigated. In both samples, the $SrTiO_3$ quantum wells are metallic and contain a high, mobile carrier density ($\sim 7 \times 10^{14} cm^{-2}$), due to interface doping from each interface [4]. In contrast to $GdTiO_3$, which is ferrimagnetic with a Curie temperature of $\sim 30$ K in bulk [15] and $\sim 20$ K in the samples with 4 nm $GdTiO_3$ [5], $SmTiO_3$ is antiferromagnetic with a Neel temperature of $\sim 50$ K [16]. Electrical and structural characterization, as well as growth details have been described elsewhere [5, 14, 17, 18]. The carrier mobility was an order of magnitude higher than in conducting, doped $GdTiO_3$ or $SmTiO_3$, showing that electrical transport occurs only in the $SrTiO_3$ quantum well (further evidence comes from magnetotransport, see below, Seebeck measurements, which are $SrTiO_3$-like [19], and the band offsets, which favor charge transfer into the $SrTiO_3$ [4, 20]).

Electrical contacts were deposited by electron beam evaporation in van der Pauw geometry with a shadow mask and consisted of 40-nm Ti/400-nm Au, with the Au contact being the topmost layer. Magnetoresistance and Hall data were collected using a Physical Property Measurement System (Quantum Design PPMS Dynacool). The system's resistivity option was used for Hall and sheet resistance and the electrical transport option for magnetoresistance. The latter utilizes an internal lock-in technique with a frequency of 70.1 Hz and an averaging time of 20 s per measurement. An external junction box allows assigning any function to any contact pad using external cables, and was used to confirm that the same behavior was measured within each possible contact geometry for a specific measurement geometry. The magnetoresistance was measured between ±1 Tesla, pausing every 0.016 T to collect data using the internal lock-in. The sweep rate between data collection points was 0.015 T/s. Multiple sweep rates were checked and no noticeable difference was detected. Longer averaging times showed no appreciable change in the resistance measurement. The angular dependence of the magnetoresistance was characterized by varying the angles $\alpha$ and $\beta$ between the current, **j**, and the magnetic field, **B**. As shown in Fig. 1, $\alpha$ lies in the film plane ($\alpha = 0°$ for **B** and **j** parallel), while $\beta$ is the angle between the film plane and **B** ($\beta = 0°$ if **B** lies in the film plane). To vary $\beta$, a horizontal rotator was used. The angle $\alpha$ was set to 0° or 90° by rotating the external leads, and to 45° by adjusting the placement of the sample on the mounting puck.

The overall longitudinal magnetoresistance ($R_{xx}$) of all samples, including SmTiO$_3$/SrTiO$_3$/SmTiO$_3$, was negative at low temperatures, independent of whether the field was in- or out-of-plane. This is contrary to what is expected for weak localization of a two-dimensional system, which is one possible origin of negative magnetoresistance. Negative magnetoresistance can, however, also be due to ferromagnetism or antiferromagnetism of the 2DEG [21, 22]. However, only the GdTiO$_3$/SrTiO$_3$/GdTiO$_3$ samples exhibited magnetoresistance hysteresis, which was superimposed on the slowly varying negative $R_{xx}$ background [see Fig. 2(a)]. The onset temperature for the hysteresis is about 5 K [23]. Thus the presence of the ferrimagnetic GdTiO$_3$ is a necessary requirement for ferromagnetism in these samples.

Figure 2 shows the relative changes of $R_{xx}$ and transverse magnetoresistance (planar Hall, $R_{xy}$), respectively, for the GdTiO$_3$/SrTiO$_3$/GdTiO$_3$ quantum well structure for three different values of $\alpha$, at $\beta = 0°$. Hysteretic behavior is observed for all $\alpha$, except for $\alpha = 0°$ in the planar Hall geometry. Both the transverse and the longitudinal hysteresis show a crossover from dip to peak behavior with $\alpha$. The relative changes are larger in $R_{xy}$, where the peak for $\alpha = 45°$ corresponds to a ~20 % change. The absolute resistance changes are, however, comparable.

To distinguish between SMR and AMR, we note that the resistivity changes as a function of magnetization orientation for the two effects are described by [8, 24, 25]:

$$\text{SMR:} \qquad \rho_{xx} = \rho_0 - \Delta\rho_S m_y^2, \qquad \rho_{xy} = \Delta\rho_S m_x m_y. \qquad (1)$$

$$\text{AMR:} \qquad \rho_{xx} = \rho_\perp + \Delta\rho_A m_x^2, \qquad \rho_{xy} = \Delta\rho_A m_x m_y. \qquad (2)$$

Here, $\rho_{xx}$ and $\rho_{xy}$ are the longitudinal and transverse resistivities, $\rho_0$ is a constant [24], and $\Delta\rho_S$ and $\Delta\rho_A$ are the resistivity changes for the SMR and AMR, respectively. $\Delta\rho_A = \rho_\parallel - \rho_\perp$, where $\rho_\parallel$ and $\rho_\perp$ are the resistivities with the field parallel and perpendicular to the current, respectively. The components of the magnetization along the $x$ and $y$ axes are $m_x = \cos\alpha\cos\beta$ and $m_y = \sin\alpha\cos\beta$. It is clear [Eq. (1)] that $R_{xx}$ for $\beta = 0°$ does not behave according to the SMR effect, because the $R_{xx}$ should be maximized for $\alpha = 0°$ and minimized for $\alpha = 90°$, as $\Delta \rho_S > 0$ always applies [24]. The experimental behavior is opposite, namely, at large fields, when the magnetization should be parallel to **B**, $R_{xx}$ decreases for $\alpha = 0°$ and increases for $\alpha = 90°$ [Fig. 2(a)]. Instead, $R_{xx}$ at $\beta = 0°$ follows AMR behavior with $\Delta \rho_A < 0$, for which the resistance should be maximized for $\alpha = 90°$ and minimized $\alpha = 0°$ [Eq. (2)], as is indeed observed. $R_{xy}$ at $\beta = 0°$ is also consistent with AMR and $\Delta \rho_A < 0$. In this case, $R_{xy}$ should be maximized (positive) for $\alpha = 135°$, minimized (negative) for $\alpha = 45°$, and zero for $\alpha = 0°$, exactly as is observed [see Fig. 2(b)].

The two effects can be further distinguished by measuring the change in resistance as the magnetization is rotated out-of-plane, at a fixed $\alpha$. Figure 3 shows the $\beta$-angle dependence of $R_{xx}$ for $\alpha = 0°$ and $\alpha = 90°$, respectively, at three different values of $B$. For AMR, $\Delta \rho_A < 0$ and $\alpha = 0°$ we expect a cosine-dependence on $\beta$, with $R_{xx}$ increasing, as $\beta$ is increased, which is observed when $B$ exceeds the demagnetization field [see Fig. 3 (a)]. In contrast, for SMR, $R_{xx}$ should be independent of $\beta$ for $\alpha = 0°$. The behavior at $\alpha = 90°$ is more complicated, because it shows a $\beta$-angle dependence [Fig. 3 (b)]. For AMR and $\alpha = 90°$, $R_{xx}$ should be independent of $\beta$. It is, however, also inconsistent with SMR, for which changing $\beta$ away from 0° should cause $R_{xx}$ to increase, not decrease, as is observed. The anisotropy in $R_{xx}$ at $\alpha = 90°$ thus has a different origin, and most likely reflects the two-dimensionality of the system, which may also contribute to the offset seen for $\alpha = 0°$ [Fig. 3(a)], i.e. the minimum of $R_{xx}$ is shifted slightly from $\beta = 0°$. The SmTiO$_3$/SrTiO$_3$/SmTiO$_3$ quantum well structure also shows a decrease in $R_{xx}$ as $\beta$ is changed from 0° to 90° for $\alpha = 90°$ (see Fig. 4), confirming that the anisotropy is independent of ferromagnetism.

The occurrence of AMR demonstrates that the SrTiO$_3$ quantum wells in the GdTiO$_3$/SrTiO$_3$/GdTiO$_3$ structures are themselves ferromagnetic. The results also provide evidence for the importance of spin-orbit coupling in these SrTiO$_3$ quantum wells, which is necessary for AMR. The ferromagnetism is a result of exchange coupling, as it does not appear in quantum wells bound by SmTiO$_3$. The origin of the negative AMR ($\Delta\rho_A < 0$) requires further theoretical investigations into the microscopic mechanisms, but there may be similarities to compressively strained, two-dimensional, magnetic III-V semiconductors, which show negative AMR [26]. We note that the SrTiO$_3$ quantum wells are under compressive strain from the LSAT substrate. The fact that the AMR is relatively weak may explain why an anomalous Hall effect could not be detected [23].

The ferromagnetic properties of the quantum well are clearly distinct from those of the GdTiO$_3$. For example, at 2K, the coercive field of the GdTiO$_3$ in these samples is about 0.02 T [5], whereas the dips/peaks in the magnetoresistance of the quantum wells appear at around 0.1 T. The onset of hysteresis in the quantum wells occurs near ~ 5 K, whereas the Curie temperature of the 4 nm GdTiO$_3$ is ~ 20 K. We also note that ferromagnetism and metallic conduction do not coexist in the rare earth titanates [27]. Octahedral distortions, a requirement for ferromagnetism in the *insulating* rare earth titanates [28], do not appear until the quantum wells are thinner than those investigated here, and insulating [14].

We hope that these results provide an incentive for future theoretical studies that examine the relative role of different parameters (exchange, orbital character of the occupied subbands in the SrTiO$_3$, and carrier density) in the magnetism, its dependence on the thickness of the quantum well, and it microscopic nature (i.e. skyrmions [29]). Because the magnetism is induced by a ferromagnetic insulator, which does not interfere with electrical transport, the structures show potential for combining correlated phenomena and engineering novel states of matter. The nearly isotropic negative magnetoresistance for the quantum wells embedded in $SmTiO_3$ suggests that proximity effects allow for inducing antiferromagnetism as well.

The authors gratefully acknowledge Gerrit Bauer, Jim Allen, Ezekiel Johnston-Halperin, and Leon Balents for very helpful discussions. The work was supported by a MURI program of the Army Research Office (Grant No. W911-NF-09-1-0398) and DARPA (Grant No. W911NF-12-1-0574). C. A. J. also received support from the National Science Foundation through a Graduate Research Fellowship. Acquisition of the oxide MBE system used in this study was made possible through an NSF MRI grant (Award No. DMR 1126455). The work made use of central facilities of the UCSB MRL, which is supported by the MRSEC Program of the National Science Foundation under Award No. DMR-1121053. The work also made use of the UCSB Nanofabrication Facility, a part of the NSF-funded NNIN network.

**Figure Captions**

**Figure 1.** Schematic of the measurement geometry, where **j** indicates the current, $\alpha$ is the in-plane angle between **B** and **j**, and $\beta$ is the out-of-plane angle. The SrTiO$_3$ quantum well is sandwiched between two GdTiO$_3$ layers, epitaxially grown on LSAT.

**Figure 2.** (a) Relative changes in the longitudinal magnetoresistance as a function of in-plane angle $\alpha$, at $\beta = 0°$. $\Delta R_{xx} = R_{xx}(B) - R_{xx}(0)$, where $R_{xx}(0)$ is the resistance at zero magnetic field. (b) Relative changes in the transverse magnetoresistance as a function of in-plane angle $\alpha$, at $\beta = 0°$. $\Delta R_{xy} = R_{xy}(B) - R_{xy}(0)$, where $R_{xx}(0)$ is the resistance at zero magnetic field. All measurements are at 2 K.

**Figure 3.** Angular out-of-plane ($\beta$) dependence of $R_{xx}$ at three different B-fields for (a) $\alpha = 0°$ and (b) $\alpha = 90°$. All measurements are at 2 K.

**Figure 4.** Relative changes in the longitudinal magnetoresistance for a SmTiO$_3$/SrTiO$_3$/SmTiO$_3$ sample for $\alpha = 90°$ and $\beta = 90°$ (bottom) and $\beta = 0°$ (top), respectively. The SmTiO$_3$ layer thicknesses are 10 nm and the SrTiO$_3$ layer thickness is 0.4 nm. The field was swept from positive B to negative B, and back. No hysteresis is observed.

**Figure 1**

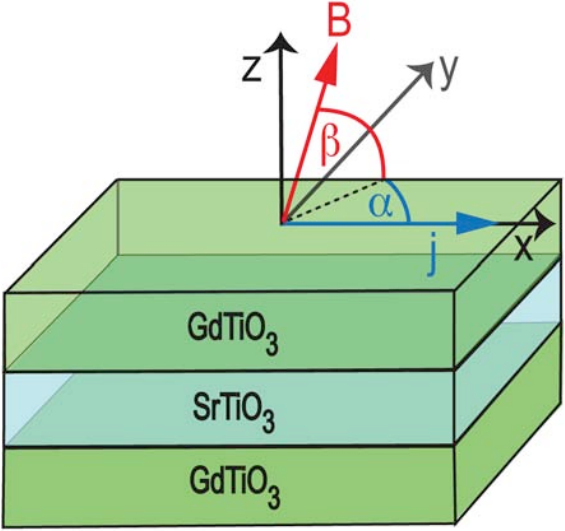

**Figure 2**

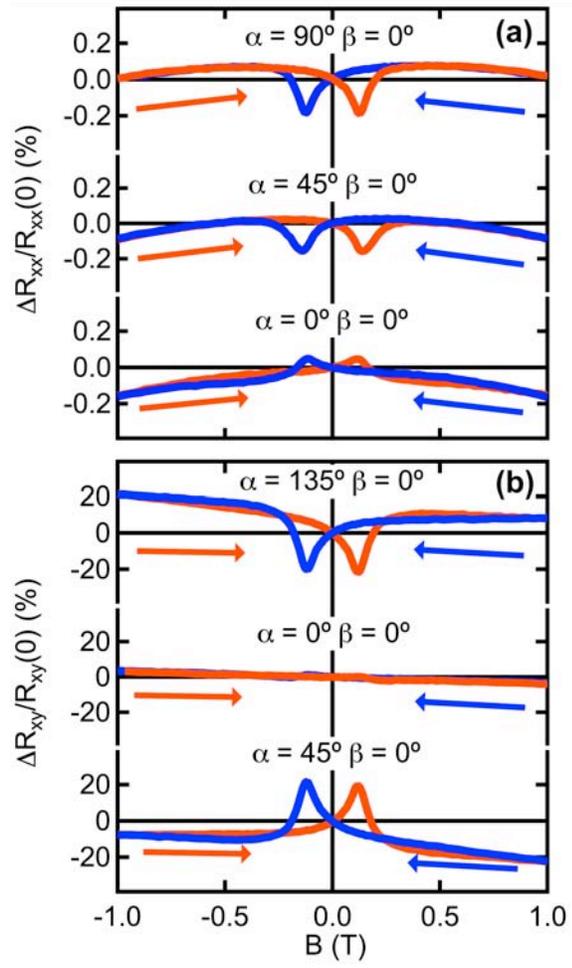

**Figure 3**

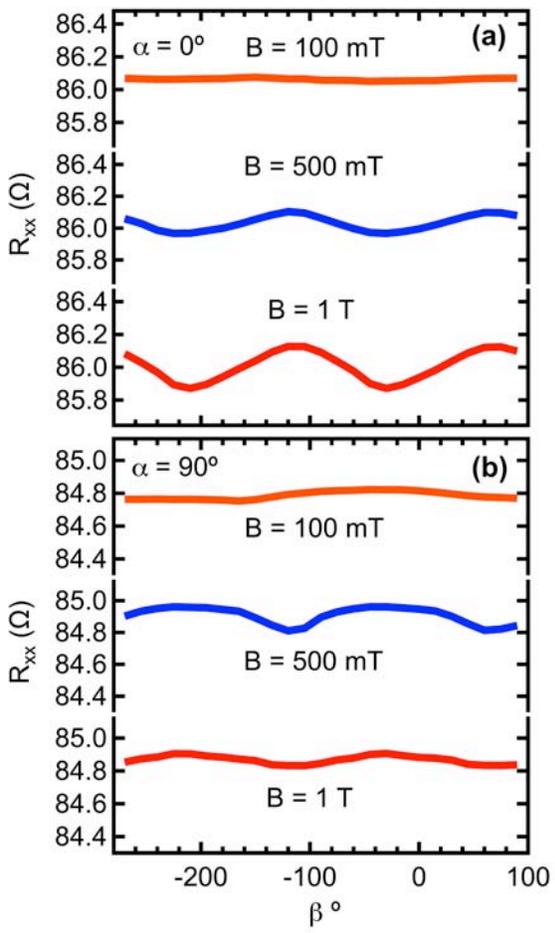

**Figure 4**

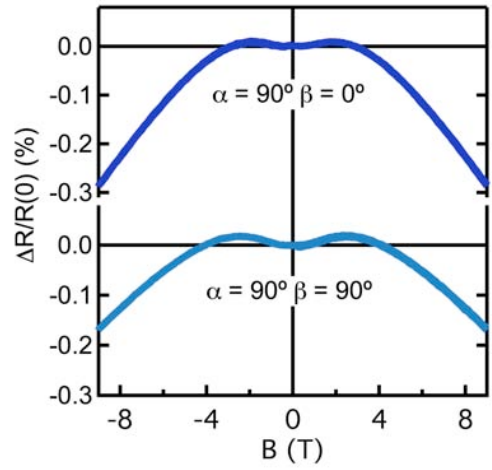

# Supplementary information
# Interface-induced magnetism in perovskite quantum wells
Clayton A. Jackson and Susanne Stemmer

**Temperature-dependence of the magenetoresistance hysteresis of the $GdTiO_3/SrTiO_3/GdTiO_3$ structures**

Figure S1 shows the longitudinal magnetoresistance for a $GdTiO_3/SrTiO_3/GdTiO_3$ sample as a function of temperature. The onset of hysteresis is about 5 K.

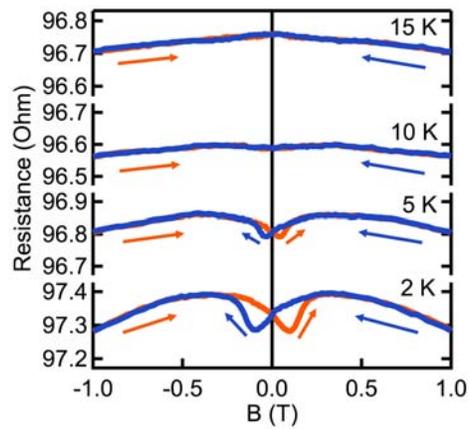

**Figure S1:** Longitudinal magnetoresistance for the $GdTiO_3/SrTiO_3/GdTiO_3$ sample for $\beta = 0°$ and $\alpha = 0°$ as a function of temperature.

**Hall resistance**

Figure S2 shows the Hall resistance ($\beta = 90°$), $R_{xy}$, for $SmTiO_3/SrTiO_3/SmTiO_3$ (left) and $GdTiO_3/SrTiO_3/GdTiO_3$ (right) quantum wells, respectively.

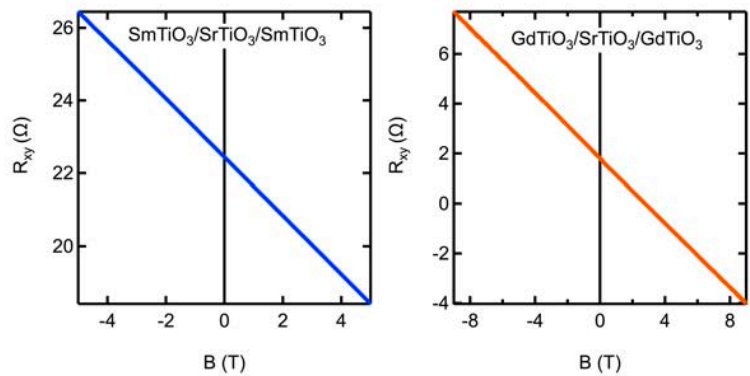

**Figure S2:** Hall resistance, $R_{xy}$, ($\beta = 90°$) for $SmTiO_3/SrTiO_3/SmTiO_3$ (left) and $GdTiO_3/SrTiO_3/GdTiO_3$ (right) samples, respectively.